\documentclass[12pt]{article}
\usepackage{amsthm,latexsym,amssymb,amsfonts,amsbsy}
\usepackage{graphicx,lscape,multibox,fancyhdr,array}
\usepackage{cite}

\def\a{\alpha}

\def\b{\beta}

\def\c{\gamma}

\def\g{\gamma}
\def\d{\delta}
\def\k{\kappa}
\def\eps{\epsilon}

\def\l{\lambda}

\def\m{\mu}
\def\n{\nu}
\def\r{\rho}

\def\s{\sigma}

\def\t{\tau}
\def\p{\pi}
\def\q{\alpha}

\def\u{\upsilon}
\def\v{\omega}
\def\w{\zeta}
\def\e{\varepsilon}

\def\be{\bar{\epsilon}}

\def\pa{\partial}

\def\ie{{\it i.e.~}}
\def\be{\begin{equation}}
\def\ee{\end{equation}}
\def\beq{\begin{eqnarray}}
\def\eeq{\end{eqnarray}}
\def\nn{\nonumber}
\def\ca{{\cal A}}
\def\cb{{\cal B}}

\def\cd{{\cal D}}
\def\ce{{\cal E}}

\def\cg{{\cal G}}

\def\ci{{\cal I}}

\def\cn{{\cal N}}
\def\co{{\cal O}}

\def\car{{\cal R}}
\def\cs{{\cal S}}
\def\ct{{\cal T}}
\def\cu{{\cal U}}
\def\cv{{\cal V}}
\def\cw{{\cal W}}

\newcommand{\bqn}{\begin{eqnarray}}\newcommand{\eqn}{\end{eqnarray}}

\begin{document}
 
\begin{flushright}
HU-EP-04/28\\
DAMTP-2004-48
\end{flushright}
\vskip 1.5cm

\begin{centering}

{\Large {\bf A classification of local Weyl invariants in $D=8$}}

\vspace{1.5cm}
{\bf Nicolas Boulanger$^{a}$ and
 Johanna Erdmenger$^{b}$}\\
\vspace{1cm}
{\small
 $^a$ Department of Applied Mathematics and Theoretical Physics\\ 
 Wilberforce Road, Cambridge CB3 0WA, UK\\
 \vspace{.2cm}
 $^b$ Institut f\"{u}r Physik\\
 Humboldt-Universit\"at zu Berlin \\ Newtonstra\ss e  15, D-12489 Berlin,
Germany
}     \\
 \vspace{.8cm}
{\small{\tt N.Boulanger@damtp.cam.ac.uk, jke@physik.hu-berlin.de}}

\vspace{.8cm}

\begin{abstract} 
 Following a purely algebraic procedure, we provide an exhaustive classification of local 
 Weyl-invariant scalar densities in dimension $D=8\,$. 
\end{abstract}

\vspace*{.2cm} 
\end{centering}

\section{Introduction}
\label{sec:Introduction}

The classification of conformal invariants is of interest both to
mathematicians and to physicists. In mathematics, local conformal invariants 
have been studied intensively for a long time. Recently, many new
results have been derived, in particular
in relation to the {\it Q-curvature}. This is a scalar in even dimensions 
which integrates to a global conformal invariant (see for instance
\cite{Branson}
and further references therein). 

Within physics, local conformal invariants are significant
in particular as contributions to the {\it Weyl anomaly} of quantum
field theories in curved space backgrounds.
In a theory which is both diffeomorphism and conformally invariant at
the classical level, the Weyl (or trace, or conformal) anomaly is the 
failure to preserve conformal invariance at the quantum 
level\footnote{Diffeomorphism invariance has to be imposed also at the
quantum level in order to ensure energy conservation. 
This is always possible.}. 
This manifests itself in anomalous contributions to the trace of the energy-momentum
tensor. These contributions involve local conformal invariants in
particular.
The Weyl anomaly was discovered more than 20 years 
ago \cite{Duff20}, but an explicit classification of all the possible Weyl anomaly terms 
in dimension $D\geq 8$ has not yet been given.

The Weyl anomaly has considerable importance as a test on
the AdS/CFT correspondence \cite{HenningsonSkenderis}: A
supergravity calculation of the Weyl anomaly which uses the
Fefferman-Graham ambient metric \cite{FeffermanGraham} yields the same
result for the anomaly as the corresponding field theory calculation
within $\cn=4\,$ super Yang-Mills theory. - For related work see
also \cite{[17]} and the recent \cite{Skenderis2+Theisen2}.

On the basis of observations coming from dimensional regularization and power 
counting, S. Deser and A. Schwimmer \cite{[16]} distinguish two
types of Weyl anomalies in arbitrary even dimensions. 
The {\textsl{type A}} anomaly is proportional to 
the Euler density of the (even-dimensional) manifold whereas the
{\textsl{type B}} anomalies are correlated 
to all strictly Weyl-invariant scalar densities involving powers of the Riemann  
tensor and its covariant derivatives. 

The Euler density of an even-dimensional oriented Riemannian manifold is well-known, 
\begin{eqnarray}
        \ce_{D=2n}\; \propto \;
        {R}_{a_1b_1}\wedge \cdots \wedge 
       {R}_{a_nb_n}\;\eps^{a_1b_1\ldots a_n b_n}\,,  \nn
\end{eqnarray}
where ${R}^a_{~b}$ is the curvature two-form.

It is the purpose of this note to communicate an exhaustive classification of the  
type B Weyl anomalies in a spacetime of dimension $D=8\,$.

In addition to the seven local invariants involving four copies of the
Weyl tensor (which have already been given in \cite{Fulling}), we find
five further local conformal invariants in $D=8$. One of them involves
four derivatives (the square of the Laplace operator), and thus is the
eight-dimensional analogue of the six-dimensional 
Parker-Rosenberg invariant \cite{FeffermanGraham, Rosenberg} (see also
\cite{Johanna}). -- The other
four invariants in $D=8$ involve two derivatives.
In other words, among the five invariants 
$\ci_j=\sqrt{-g}\,I_j\,$, $j=1,\ldots ,5\,$, only the first one starts with a 
term quadratic  in the Weyl tensor, whereas the other four have a leading-order term which 
is cubic in the (covariant derivatives of the) Weyl tensor. Schematically,  

\begin{eqnarray}
        I_1\,&\sim&\, C^{\a\b\c\d}\Box^2C_{\a\b\c\d}+({\rm{more}})\,,
\nonumber \\
        I_k \,&\sim&\,  C \cdot (\nabla C)\cdot(\nabla C) + C \cdot C \cdot(\nabla\nabla C) 
              + ({\rm{more}}) \,,\quad k=1,2,3,4\,. \nonumber 
\end{eqnarray}

Within the AdS/CFT correspondence, one particular global Weyl invariant
in $D=8$ has been calculated  using the Dirichlet obstruction in 
\cite{Alvarez}.\footnote{A discussion of the holographic anomaly in $D=8$ has 
also been given in \cite{Odintsov}.} This invariant differs from the local ones 
obtained here by total derivatives.
\vspace*{.2cm}


Our method is purely algebraic. 
First, we write down a basis of dimension eight scalars which is suited to the 
computation of Weyl invariants. 
A crucial ingredient in this construction is what we call the  
{\it Weyl-covariant derivative} ${\cal D}\,$.
This derivative is defined on the Weyl tensor and obeys the nice property that 
its commutator $[{\cal D}, {\cal D}]$ involves the Weyl tensor $C^{\a}_{~\,\b\c\d}$ 
and the Cotton tensor $\tilde C_{\beta \gamma \delta}\,$. 
This commutation relation justifies the name of $\,\cd$ and makes a link with 
conformal geometry.
Indeed, a metric on a manifold of dimension $D\geq 4$ is conformally flat if and only 
if its Weyl tensor identically vanishes. In  $D\geq 4$, this condition 
implies that the Cotton tensor is also zero.
On the other hand, in $D=3\,$, the Weyl tensor vanishes identically 
whereas the Cotton tensor does not in general, and a metric on a manifold of dimension $D=3$ is conformally flat 
if and only if its Cotton tensor identically vanishes.

The task of constructing the basis of dimension eight scalars 
can be rephrased into a problem of  
representation theory for the orthogonal group in eight dimensions. 
The dimension of our basis is eighteen. This corresponds to the 
multiplicity of the trivial one-dimensional representation in the decomposition of 
the tensor products corresponding to the different products of (covariant 
derivatives of) Weyl tensors. Group representation theory also determines
the structure of our eighteen terms.

After constructing the basis, we calculate its Weyl variation. This is simplified by the property
that the Weyl transformation of the tensors in the basis bring in only one derivative of 
the Weyl parameter.   
Setting the variation to zero leads to conditions on the eighteen coefficients in the 
original scalar, which we solve. 

It is crucial to keep track of potential redundancies due to the Bianchi identities 
during the entire calculation.
Whenever the $\ell\,$th covariant derivative of the Riemann tensor appears in 
the variation of the basis, we expand it in irreducible representations of 
$GL(8)\,$. 
This expansion enables us to identify the Bianchi identities and to treat them 
appropriately.  

We resort to algebraic programming with the programs {\textsl{FORM}} \cite{vermaseren}, 
{\textsl{Ricci}} \cite{Lee} (package for {\textsl{Mathematica}}) and 
{\sffamily{LiE}} \cite{LiE}. 
This allows to cope with the volume of the computation and to perform several 
group theoretical calculations. 
When manipulating Young tableaux, {\textsl{FORM}} and {\textsl{Ricci}} 
assist us in obtaining the exact tensor expressions and the Clebsch-Gordan 
coefficients in the expansion of the $\ell\,$th covariant derivative of the Riemann 
tensor in terms of irreducible representations of $GL(8)\,$.   
The program {\sffamily{LiE}} proves to be particularly useful when decomposing  
covariant derivatives of Weyl tensors in irreducible representations of 
$O(8)\,$.


\section{Classification Result}
\label{sec:ClassificationResult}

After some elementary definitions related to Riemann and Weyl invariants 
(Section \ref{Definitions}), we introduce  the notion of {\textsl{jet 
space}} in Section \ref{jet}. Using this concept, we explain in Section \ref{tensors} the method that we 
follow in order to construct all Weyl-invariant scalar densities in $D=8\,$.  
A well-suited basis for the computation of Weyl-invariant scalar densities is then 
presented in  Section \ref{basis}. Finally, our main result is stated in Section 
\ref{theo}. 

\subsection{Definitions}
\label{Definitions}

The Weyl transformation of the metric is defined as follows :
\be
\d_{\s} g_{\a\b}=2 \,\s \, g_{\a\b}\,,
\label{weyltransfo}
\ee
where $\s(x)$ is the Weyl parameter. 
The Weyl tensor $C^{\a}_{~\,\b\g\d}$ in dimension $D$ is defined by
\begin{eqnarray}
C^{\a}_{~\,\b\g\d}&=&R^{\a}_{~\,\b\g\d}-2\left(
\d^{\a}_{\,[\g}K_{\d]\b}-g_{\b[\g}K_{\d]}^{~\;\a} \right),
\label{weyl2} \\
        K_{\a\b}&=&\frac{1}{D-2}(R_{\a\b}-\frac{1}{2(D-1)}g_{\a\b}R )\,,
\end{eqnarray}
where $R^{\a}_{~\,\b\g\d}\,$, $R_{\a\b}$ and $R$ are the Riemann tensor, 
the Ricci tensor and the  scalar curvature, respectively. 
Square brackets denote antisymmetrization with strength one, whereas 
symmetrization 
is denoted by round brackets. 
The Weyl tensor is strictly Weyl invariant, 
\begin{eqnarray}
        \d_{\s}C^{\l}_{\,~\m\n\r} = 0\,,\quad\forall\; D\,.\nn
\end{eqnarray}

\vspace{.2cm}
A Riemann-invariant scalar is a polynomial in the variables of the space $\cg$
\begin{eqnarray}
        \cg &=& \{ \pa_{\a_1}\ldots\pa_{\a_k} g_{\m\n}\,,\quad k\in {\mathbb{N}} \}\nn
\end{eqnarray}
and\footnote{Although our treatment is compatible with a metric of signature $(p,q)$,  $p+q=D$, we choose a Lorentzian signature for definiteness and convenience of notations. 
The metric is supposed to be invertible.} 
$(\sqrt{-g})^{-1}\equiv(-{\rm{det}}\, g_{\m\n})^{-\frac{1}{2}}\,$, 
which is independent of the coordinate system. 
It is a standard result that one can form the Riemann-invariant scalars by taking 
polynomial scalar functions in $(\sqrt{-g})^{-1}$ and the variables of a smaller space 
$\ct\subset \cg$
\begin{eqnarray}
\ct &=& \{ g_{\m\n}\,,\nabla_{\a_1}\ldots\nabla_{\a_k}R^{\m}_{~\,\n\r\s}\,, 
\quad k=0,1,\ldots\;\}\,,\nn
\end{eqnarray}
where $\nabla$ is the covariant derivative associated with the Levi-Civita connection.


The Riemann-invariant scalar densities are obtained by multiplying 
$(\sqrt{-g})$ by products of covariant derivatives of the Riemann tensor with the metric $g_{\m\n}$, the inverse metric $g^{\a\b}$ and/or the completely antisymmetric tensor $\epsilon^{\m_1\ldots\m_D}\equiv\frac{1}{\sqrt{-g}}\e^{\m_1\ldots\m_D}$, and then by contracting all the indices. Roughly, 
\begin{eqnarray}
        \ca = \sqrt{-g}\;{\rm{tr}} \Big(\epsilon\otimes \cdots\otimes \epsilon \otimes 
        g\otimes \cdots \otimes g \otimes g^{-1}\otimes \cdots \otimes g^{-1} \otimes
        \nabla^{\ell_1}R\otimes\cdots\otimes\nabla^{\ell_r}R\Big)\nonumber \,,
\end{eqnarray}
where $\nabla^{\ell_1}R$ denotes a monomial of degree one in the Riemann tensor and 
of degree $\ell_1$ in the covariant derivative.
\vspace{.2cm}

We name $\ci\;$ a {\textit{Weyl-invariant scalar density}} if it is a Riemann-invariant 
scalar density which is {\textit{strictly}} invariant under the Weyl
transformation (\ref{weyltransfo}): 
\begin{eqnarray}
        \d_{\s} \ci = 0\,.\nn
\end{eqnarray}
That $\ci$ must be a scalar {\textit{density}} in order to satisfy the above 
equation is easily seen by inspection of the Weyl transformation of the Riemann tensor and 
 $\sqrt{-g}\,$. 
In what follows, we will not always specify that $\ci$ must
be a scalar density, being understood that it suffices to multiply a scalar $I$ by
 $\sqrt{-g}$ to obtain a density $\ci\,$.   

\subsection{Jet space}
\label{jet}

We consider the fields $\phi^i\equiv \{g_{\m\n}\,,\s\}$ 
and their partial derivatives of first and higher order as independent local 
coordinates of a so-called jet space\footnote{For rigorous definitions of jet spaces, 
see e.g. \cite{jet}.}. 

The quantities that we manipulate (tensors, Riemann-invariant scalars {\textit{etc}}) 
will be 
seen as functions of the field variables $\phi^i$ and a finite number of their derivatives, 
$f=f([\phi])$, where the notation $[\phi]$ means dependence on 
$\phi^i,\phi^i_{\mu},\dots, \phi^i_{(\mu_1\dots\mu_k)}$
for some finite $k$ (with $\phi^i_{\mu}\equiv \pa_\mu\phi^i$,
$\phi^i_{(\mu_1\dots\mu_k)}\equiv \pa_{\mu_1}\dots \pa_{\mu_k}\phi^i\,$).
A local function $f([\phi])$ is a function on the ``jet space of order $k$" 
$J^k$ (for some $k$).

The fields and their various derivatives are considered completely independent 
coordinates in $J^k$ except that the various derivatives commute, so that only the completely 
symmetric combinations are independent coordinates.  


\subsection{Weyl-covariant tensors}
\label{tensors}

As mentioned earlier, it is sufficient to restrict the computation 
of Riemann-invariant scalars to the space $\ct$ of tensors under 
diffeomorphisms,
\begin{eqnarray}
        \{\ct^i\} = \{ g_{\m\n}\,,\nabla_{\a_1}\ldots\nabla_{\a_k}R^{\m}_{~\,\n\r\s}\,, 
\quad k=0,1,\ldots\;\}\,.\nn
\end{eqnarray}
Similarly, one can simplify the classification of the Weyl-invariant scalars by 
considering scalar functions in a smaller space $\cw\subset\ct\,$.  

In this section, we exhibit a local, invertible change of jet coordinates in $J^6$ 
from the set   
\begin{eqnarray}
\cs=\{g_{\m\n}\,,\s\,,\s_{\m}\,,\nabla_{\a_1}\ldots\nabla_{\a_k}
R^{\m}_{~\,\n\r\l}\,,\nabla_{(\a_1}\ldots\nabla_{\a_{k+2})}\s\,,\quad k=0,\ldots,4\}\nn         
\end{eqnarray}
to local coordinates $\cb=\{\cu^{\ell},\cv^{\ell},\cw^i\}$ satisfying 
\begin{eqnarray}
        \d_{\s} \cu^{\ell} &=& \cv^{\ell}\quad \forall ~\ell\,,
        \label{cond1}\\
        \d_{\s} \cw^{i} &=& \car^i(\cw)\quad \forall ~i\,,
        \label{cond2}
\end{eqnarray}
and such that any function $f(\ct^i)$ can be expressed as a function 
$\tilde{f}(\cu^{\ell},\cw^i)$ of the new variables $\cu^{\ell}$ and $\cw^i\,$. 

Then, computing the Weyl transformation of a scalar function 
$f(\ct^i)=\tilde{f}(\cu^{\ell},\cw^i)\,$ and demanding that the 
result should vanish,  
\begin{eqnarray}
        0=\d_{\s}\tilde{f}(\cu^{\ell},\cw^i)=
        \sum_i \frac{\pa \tilde{f}}{\pa \cw^i}{\cal R}^i(\cw) + 
        \sum_{\ell} \frac{\pa \tilde{f}}{\pa
        \cu^{\ell}}\;\cv^{\ell}\; , \nn 
\end{eqnarray}
implies the conditions $\frac{\pa \tilde{f}}{\pa \cu^{\ell}}=0\;\;\forall \ell\,$, 
since the variables $\cv^{\ell}$ are independent coordinates in $J^6\,$.
We finally have to solve the remaining conditions 
$\sum_i \frac{\pa \tilde{f}}{\pa \cw^i}{\cal R}^i(\cw)=0\,$.
\vspace*{.2cm}

{\textit{Consequently, in order to classify the Weyl invariants, it is sufficient to
 restrict oneself to scalar functions $\tilde{f}(\cw^i)$ of the sole variables $\cw^i$.}} 
\vspace*{.2cm}

\noindent To proceed, we state the following lemma:

\newpage
{\bfseries Lemma [Decomposition of the jet of the Riemann tensor]}
\newline\noindent 
{\textit{Every function of the Riemann tensor and its covariant derivatives 
can be  written as a function of the Weyl tensor and its covariant 
derivatives plus the completely symmetric tensors 
$\nabla_{(\l_1\l_2\ldots\l_k}K_{\a\b)}\,$.}}
\vspace*{.4cm}

\noindent Diagrammatically, this Lemma translates to
\vspace*{.5cm}

{\footnotesize
\begin{center}
  \begin{picture}(22,5)(3,5)
\multiframe(10,3)(10.5,0){1}(10,10){}
\end{picture}  
$\bigotimes$
  \begin{picture}(15,5)(9,5)
\multiframe(10,3)(10.5,0){1}(10,10){}
\end{picture} 
$\bigotimes\ldots$
$\bigotimes$
  \begin{picture}(15,5)(9,5)
\multiframe(10,3)(10.5,0){1}(10,10){}
\end{picture} 
$\bigotimes$
  \begin{picture}(32,0)(9,0)
\multiframe(10,3)(10.5,0){2}(10,10){}{}
\multiframe(10,-7.5)(10.5,0){2}(10,10) {}{}
\end{picture} 
$\simeq$
  \begin{picture}(22,5)(3,5)
\multiframe(10,3)(10.5,0){1}(10,10){}
\end{picture}  
$\bigotimes$
  \begin{picture}(15,5)(9,5)
\multiframe(10,3)(10.5,0){1}(10,10){}
\end{picture} 
$\bigotimes\ldots$
$\bigotimes$
  \begin{picture}(15,5)(9,5)
\multiframe(10,3)(10.5,0){1}(10,10){}
\end{picture} 
$\bigotimes$ 
\begin{picture}(35,0)(9,0)
\multiframe(10,3)(10.5,0){2}(10,10){}{}
\multiframe(10,-7.5)(10.5,0){2}(10,10) {}{}
\put(33,-10){t.f.}
\end{picture}
$\bigoplus$
  \begin{picture}(22,5)(3,5)
\multiframe(10,3)(10.5,0){2}(10,10){}{}
\multiframe(31,3)(10.5,0){1}(19,10){$\ldots$}
\multiframe(50,3)(10.5,0){1}(10,10){}
\put(-355,-10)
{$\underbrace{\hspace*{103pt}}_{\mbox{\scriptsize{k covariant derivatives}}}$}
\put(-230,-18){$R_{\a\b\g\d}$}
\put(-175,-10)
{$\underbrace{\hspace*{103pt}}_{\mbox{\scriptsize{k covariant derivatives}}}$}
\put(-45,-18){$C_{\a\b\g\d}$}
\put(3,-18){$\nabla_{(\l_1\l_2\ldots\l_k}K_{\a\b)}$}
\end{picture} {~~~~~~~~~~~~~.}
\end{center}}
\vspace*{1cm}

\noindent This result was used at the linearized level in the 
BRST-cohomological derivation of Weyl 
gravity in \cite{Boulanger:2001he}; a proof is given in appendix \ref{jetspace}.
\vspace*{.2cm}

With this decomposition, we have almost achieved the change of jet coordinates from 
$\cs$ to $\cb\,$ because the Weyl transformation of the covariant derivatives  
$\{\nabla_{(\l_1}\ldots\nabla_{\l_k}K_{\a\b)}\,,\;k=0,\ldots, 4\}\,$ yields the covariant 
derivatives $\{\nabla_{(\l_1}\ldots\nabla_{\l_{k+2})}\s\,,\;k=0,\ldots, 4\}\,$, up to lower 
order terms in the variables of $J^6\,$. 
First, one has $\d_{\s}K_{\a\b}=-\nabla_{\a}\nabla_{\b}\s\,$.
Then, performing the invertible change of coordinate in $J^6$,
\begin{eqnarray}
\nabla_{(\mu}\nabla_{\a}\nabla_{\b)}\tilde{\s}\equiv
-\nabla_{(\mu}\nabla_{\a}\nabla_{\b)}\s + 
4 K_{(\a\b}\nabla_{\m )}\s -2 g_{(\m\a}K_{\b)\l}\nabla^{\l}\s \,,\nonumber
\end{eqnarray}
we have
\begin{eqnarray}
 \d_{\s} \nabla_{(\l}K_{\a\b)} = \nabla_{(\l}\nabla_{\a}\nabla_{\b)}\tilde{\s}\,.\nonumber      
\end{eqnarray}
It is not difficult to see that one can perform similar local, invertible
 changes of coordinates 
$\nabla_{(\a_1}\ldots\nabla_{\a_{k+4})}\s\rightarrow$ 
$\nabla_{(\a_1}\ldots\nabla_{\a_{k+4})}\tilde{\s}=
-\nabla_{(\a_1}\ldots\nabla_{\a_{k+4})}\s+\co(J^{k+3})\,$, 
$k=0,1,2\,$, such that 
\begin{eqnarray}
\d_{\s}\nabla_{(\a_1}\ldots\nabla_{\a_{k+2}}K_{\a_{k+3}\a_{k+4})}=
\nabla_{(\a_1}\ldots\nabla_{\a_{k+4})}\tilde{\s}\nn\,.  
\end{eqnarray}

The next (and more difficult) task is to find a local, invertible change of coordinates in the sector of the covariant derivatives of the Weyl tensor,
\begin{eqnarray}
        \nabla_{\a_1}\ldots \nabla_{\a_{k}}C^{\b}_{~\g\d\eps}\rightarrow
        \cd_{\a_1}\ldots \cd_{\a_{k}}C^{\b}_{~\g\d\eps}= 
        \nabla_{\a_1}\ldots
        \nabla_{\a_{k}}C^{\b}_{~\g\d\eps}+\co(J^{k+1})\,,~\;k=0,\ldots,4
        \; ,
\nn
\end{eqnarray}
where the variation of the new jet coordinates 
$\cd_{\a_1}\ldots \cd_{\a_{k}}C^{\b}_{~\g\d\eps}$ does not bring in more than one covariant 
derivative of $\s\,$. In other words, the variation of these new variables should not mix    
with the $\cv^{\ell}\,$-variables, which would be in conflict with the condition 
(\ref{cond2}). 

At order zero and one in the covariant derivative of the Weyl tensor, the property 
(\ref{cond2}) is already satisfied. It is possible to satisfy the same condition at 
the next orders.  
Denoting $W^{\Delta_0}\equiv C^{\b}_{~\,\g\d\eps}\,$, 
 $W^{\Delta_1}\equiv \cd_{\a_1}C^{\b}_{~\,\g\d\eps}\,$, $W^{\Delta_2}\equiv 
 \cd_{\a_2}\cd_{\a_1}C^{\b}_{~\,\g\d\eps}\,$ \textit{etc}, we have 
\begin{eqnarray}
        W^{\Delta_{1}} &\equiv& \cd_{\a_1}C^{\b}_{~\,\g\d\eps} = 
        \nabla_{\a_1}C^{\b}_{~\,\g\d\eps}\,,
        \nn \\
        W^{\Delta_{i}} &\equiv& \cd_{\a_{i}}W^{\Delta_{i-1}} = 
        \nabla_{\a_{i}}W^{\Delta_{i-1}} + 
        K_{\a_{i}\m}{\big[{\mathbf{T}}^{\m}\big]}^{\Delta_{i-1}}_{~~\Delta_{i-2}}  
  W^{\Delta_{i-2}}\,, \quad i=2,3,4\ldots \; \,
\nonumber
\end{eqnarray}
where we use Einstein's summation convention for the indices $\Delta_i$ 
$(i\in \mathbb{N})\,$. [The matrices ${\big[{\mathbf{T}}^{\m}\big]}^{\Delta_{i+1}}_{~~\Delta_{i}}\,$, 
$i=0,1,2,3\;$ are given in Appendix \ref{T}.] 
In terms of the matrices ${\big[{\mathbf{T}}^{\m}\big]}$'s, we have the following 
realization of Condition (\ref{cond2}) :  
\begin{eqnarray}
        \d_{\s} W^{\Delta_{i+1}} = \s_{\m}\; 
{\big[{\mathbf{T}}^{\m}\big]}^{\Delta_{i+1}}_{~~\Delta_{i}}\;W^{\Delta_{i}}\,,
        \label{variaW}
\end{eqnarray}
where only the first derivative of the Weyl parameter is involved.

In summary, we have 
\bqn
\{\cw^i\}      &=&  \left\{ g_{\a\b}\,,\; \s\,,\; \s_{\m}\,,\; 
\cd_{\a_1}\ldots\cd_{\a_k}C^{\b}_{~\g\d\eps}\,,~\;k=0,\ldots,4\right\}\,,
\nonumber \\
\{\cu^{\ell}\} &=&  \left\{
       \nabla_{(\l_1} \ldots \nabla_{\l_k}K_{\a\b)}\,,~\;k=0,\ldots,4 \right\}\,,
\nn \\  
\{\cv^{\ell}\}&=&\left\{ 
\nabla_{(\a_1}\ldots\nabla_{\a_{k+2})}\tilde{\s}\,,~\; k=0,\ldots,4 \right\}\,. 
\nonumber \label{set}
\eqn
\vspace*{.1cm}

{\bfseries{Remark}} : The name \emph{Weyl-covariant derivative}
for the operator $\cd$ is in agreement with the fact
that the commutator $[\cd,\cd]$ is given by the 
Weyl tensor $C^{\b}_{~\g\d\eps}$ plus the Cotton tensor $\tilde{C}_{\g\d\eps}$. 
Schematically,
\begin{eqnarray}
        {[} \cd,\cd {]} = C + \tilde{C}\,.\nn
\end{eqnarray}
 More precisely, the Weyl-covariant derivative $\cd_{\a}$ is defined on the 
 $W^{\Delta_i}$'s, $i\in \mathbb{N}\,$, and is such that
\begin{eqnarray}
  \big[\cd_{\a_{2}},\cd_{\a_1}\big]W^{\Delta_{0}} &=& 
        \big[\nabla_{\a_{2}},\nabla_{\a_1}\big]_{R\rightarrow C} W^{\Delta_{0}}\,,
        \nn \\
        \big[\cd_{\a_{i+1}},\cd_{\a_i}\big]W^{\Delta_{i-1}} &=& 
        \big[\nabla_{\a_{i+1}},\nabla_{\a_i}\big]_{R\rightarrow C} W^{\Delta_{i-1}} -  
        \tilde{C}_{\m\a_{i+1}\a_i}
        {\big[{\mathbf{T}}^{\m}\big]}^{\Delta_{i-1}}_{~~\Delta_{i-2}}W^{\Delta_{i-2}}
        \,,\quad i\geq 2\,.\nn
\end{eqnarray}
 The first term in the right-hand-side of both equations means that one applies 
 the well-known rule for the commutator of two covariant derivatives but where  
 {\emph{every Riemann tensor must be replaced by a Weyl tensor}}. 
   
It is worth pointing out that although the representation matrices 
${\big[{\mathbf{T}}^{\m}\big]}^{\Delta_{i+1}}_{~~\Delta_{i}}$
depend on the spacetime coordinates through the metric and it inverse, they are inert under 
Weyl transformations because they are built out of an equal number of 
$g_{\m\n}$'s and $g^{\a\b}$'s. 
The $\big[{\mathbf{T}}^{\m}\big]$'s 
completely determine the space $\{\cw^i\}$ of tensors which transform covariantly 
under Weyl transformations. These matrices and their properties were obtained 
following a BRST-cohomological construction along the lines of  
\cite{brandt}.
More details on the Weyl-covariant operator $\cd$ and its underlying covariant 
algebra will be given elsewhere \cite{Nicolas}.

\subsection{A basis for Weyl-invariant scalars}
\label{basis}

Now that we have at our disposal a set of Weyl-covariant tensors $\{W^{\Delta_i}\}$ 
transforming according to the formula (\ref{variaW}), we have to contract them and 
construct the most general Weyl-invariant scalar Ansatz $A_8\,$. After that, we 
must multiply $A_8$ by $\sqrt{-g}\,$, take its Weyl-transformation and then follow the 
procedure explained after  (\ref{cond1}) and (\ref{cond2}).

\vspace*{.2cm}
The construction of $A_8$ is done by using group theoretical methods. 
As explained in details in the very clear and pedagogical  \cite{Fulling}, 
the $\ell\,$th covariant derivative of the Riemann tensor 
\begin{eqnarray}
        \nabla^{\ell}R=\{R_{\a_1\a_2\a_3\a_4;\a_5\ldots\a_{\ell}}\}\,,\quad
        \ell\geq 0 \; , \nn
\end{eqnarray}
corresponds to the irreducible representation $(\ell+2,2)$ of the general linear group  
$GL(D)$ associated with the Young tableau having $\ell+2$ boxes in 
the first row and two boxes in the second row,
\begin{center}
\begin{picture}(0,40)(30,-15)
\put(-70,8){$(\ell+2,2)\sim$}
\multiframe(0,10)(10.5,0){1}(10,10){}
\multiframe(10.5,10)(10.5,0){1}(10,10){}
\multiframe(21,10)(10.5,0){1}(10,10){}
\multiframe(31.5,10)(10.5,0){1}(10,10){}
\multiframe(42,10)(10.5,0){1}(17,10){}
\multiframe(59.5,10)(10.5,0){1}(10,10){}
\multiframe(0,-0.5)(10.5,0){1}(10,10){}
\multiframe(10.5,-0.5)(10.5,0){1}(10,10){}
\put(80,9){$\cdot$}
\put(43.5,14){$\ldots$}
\put(22,9){$\underbrace{\hspace*{47pt}}_{\ell}$}
\end{picture}
\end{center}
To know the number of scalars one can built from polynomials in the Riemann tensor and its 
covariant derivatives amounts to looking at the multiplicity of the trivial one-dimensional irreducible 
representation of $GL(D)$ in the tensor product representing the corresponding polynomials. 
 
However, we must turn to the representation theory of the orthogonal
group in dimension $D=8\,$, since we are essentially dealing with the covariant 
derivatives of the 
{\textit{Weyl}} tensor. As far as the group theoretical problem of constructing a basis of 
locally independent scalars in the tensors $W^{\Delta_i}$ is concerned, 
we may replace the Weyl-covariant derivatives with the standard
covariant derivatives when considering  scalar polynomials in the covariant derivatives of 
the Weyl tensor. The difference between $\nabla_{\a_1}\ldots\nabla_{\a_i}C^{\b}_{~\,\g\d\eps}$
 and $\cd_{\a_1}\ldots\cd_{\a_i}C^{\b}_{~\,\g\d\eps}$ consists of (products of) lower-order terms in the derivatives of the metric, \ie terms which belong to $J^{i+1}\,$. 
Once the basis is found, we just have to replace every $\nabla_{\a_1}\ldots\nabla_{\a_i}C^{\b}_{~\,\g\d\eps}$ 
by $\cd_{\a_1}\ldots\cd_{\a_i}C^{\b}_{~\,\g\d\eps}\,$.


In this way we obtain a basis of dimension eight scalars of the form
\footnote{Strictly speaking, we should add 
to $A_8$ all the linearly independent quartic terms in the undifferentiated Weyl tensor. 
However, since they are trivially  Weyl invariant (after multiplication by $\sqrt{-g}$), 
they decouple from the problem. Accordingly, we do not consider them.}
\begin{eqnarray}
 A_8 &=& 
  a_1 \,C_{\r\g\m\s}\cd^{\a}\cd_{\a}\cd^{\b}\cd_{\b}C^{\r \g \m \s}
   +
    \nonumber \\ &&  
  b_1 \,\cd_{\b}C_{~ \g\m\a}^{\b}\cd_{\n}\cd^{\n}\cd_{\r}C^{\r\g\m\a} + 
    b_2 \,\cd_{\a}C_{\m\b\g\n} \cd_{\r}\cd^{\r}\cd^{\a}C^{\m\b\g\n} + 
    \nonumber \\
  &&c_1 \,\cd^{\a}\cd^{\b}C_{\g\a\b\m} \cd_{\n}\cd_{\r}C^{\g\n\r\m} + 
    c_2 \,\cd^{\g}\cd_{\g}C_{\a\m\n\b}  \cd^{\r}\cd_{\r}C^{\a\m\n\b} + 
    \nonumber \\
  &&c_3 \,\cd_{\a}\cd_{\b}C_{\n\g\m\r}\cd^{\a}\cd^{\b}C^{\n\g\m\r} + 
    c_4 \,\cd_{\a}\cd_{\g}C_{~\m\n\b}^{\g} \cd^{\a}\cd_{\r}C^{\r\m\n\b} +
    \nonumber \\
  &&d_1 \,\cd^{\r}\cd^{\s}C_{~\r\s}^{\b~~\n}C_{\b}^{~ \a\g \m} C_{\n\a\g\m} + 
    d_2 \,\cd^{\b}\cd^{\g}C_{~\,\n\r}^{\m~~\s}C_{\m\b\g\a}C_{\s}^{~\n\r\a} + 
   \nonumber \\ 
  &&d_3 \,\cd_{\s}\cd^{\s}C_{~ \g~ \m}^{\n ~ \r}C_{\a ~\b}^{~\g~\m}C_{~ \n~ \r}^{\a~\b} + 
    d_4 \,\cd^{\s}\cd_{\s}C^{\b\n\m\r}C_{\b\n}^{~~\;\a\g}C_{\a\g\m\r} + 
   \nonumber \\
  &&e_1 \,C_{~\a ~ \b}^{\g ~ \m}\cd^{\r}C_{\r\g~\m}^{~~\n}
    \cd^{\s}C_{\s~ \n~}^{~ \a~\b} + 
    e_2 \,C_{\a ~ \b}^{~\g ~\m}\cd^{\r}C_{\r\g\s\n}\cd^{\a}C_{~\m}^{\b~ \s\n} + 
    \nonumber \\
  &&e_3 \,C_{\a\b\g\m}\cd_{\n}C^{\a\b\r\s}\cd^{\n}C_{~~\;\r\s}^{\g\m} + 
    e_4 \,C_{\a~\b}^{~\g~ \m}\cd^{\r}C_{\r~~\,\m}^{~\n\s}
    \cd^{\a}C_{~\;\n\s\g}^{\b} + 
    \nonumber \\
  &&e_5 \,C_{\a\b\g\m}\cd^{\r}C_{\r\n}^{\,~~\a\b}\cd^{\s}C_{\s}^{~\,\n\g\m} + 
    e_6 \,C^{\a}{}_{\g\m\b}\cd_{\a}C_{\g\n\r\s}\cd^{\b}C_{\,~\n\r\s}^{\m} + 
    \nonumber \\  
  &&e_7 \,C_{~\m~ \b}^{\a~\g}\cd^{\n}C_{~\a~ \g}^{\r~\s}\cd_{\n}
  C_{~ \r~ \s}^{\m~\b}\,,
\label{ansatz8}
\end{eqnarray}
where the first term belongs to $J^6$ because of the four derivatives which act on a Weyl 
tensor. The two terms with coefficients $b_1$ and $b_2$ belong to $J^5\,$. The terms multiplied 
by the $c$'s and $d$'s live in $J^4$ and finally, the scalars with the $e\,$-coefficients belong to 
$J^3\,$. 

The general structure of $A_8$ is obvious for dimensional reasons. However, it is not 
necessary to resort to dimensional analysis. It suffices to use the property that the 
$W^{\Delta_i}$'s are $(1,i+3)$-type tensors which transform with one derivative of $\s\,$. 
Indeed, let $M(W)$ be a monomial of degree $r$ in the $W^{\Delta_i}$'s.
It is a $(r,s)$-type tensor with $s>r\,$. If no $\e\,$-tensor is 
to be used when contracting the $(r+s)$ indices, one must have $s-r=2n\,$, 
$n\in\mathbb{N}\,$, to 
make a scalar. One then contracts $\sqrt{-g}\,M(W)$ with a product of $p$ 
inverse metrics and 
$q$ metrics, $p-q=n\,$, to form a scalar density $\ca_8\,$. 
In the Weyl transformation of 
$\ca_8=[\sqrt{-g}\,g\cdots g\, g^{-1}\cdots g^{-1}]M(W)\,$, 
the factor between square brackets 
will be multiplied by $(D-2n)\s$ whereas $\d_{\s}M(W)=\s_{\m}(\cdots)^{\m}\,$.
Since $\s$ and $\s_{\m}$ are independent, one must require $2n=D$ if one wants that 
$\d_{\s}\ca_8$ vanishes.
The monomials we used in  (\ref{ansatz8}) are the only ones with $s-r=2n=8\,$.

To show that no parity-odd scalars can be formed, or, in other words, that no $\e\,$-tensor 
can be used to contract the indices, we must consider that each $W^{\Delta_i}$ transforms in 
the $(2+i,2\,)$-irreducible representation of $GL(8)\,$ (see e.g.~\cite{Fulling}). 
Then, if one\footnote{A product of two $\e\,$-tensors can be written in terms of the
 metric.} $\e\,$-tensor is used to contract some indices of a tensor-valued 
monomial $M(W)\,$, 
one sees that at least one tensor $W^{\Delta_i}$ from $M(W)$ must have three indices 
contracted with the completely antisymmetric tensor $\e\,$. 
Due to the symmetries of $W^{\Delta_i}\,$, it gives zero. 

\vspace*{.2cm}\noindent
{\bfseries{Construction of $A_8$}}
\vspace*{.1cm}

\noindent
Using the program {\sffamily LiE} \cite{LiE}, 
we expand every $\ell\,$th covariant derivative of the Weyl tensor in irreps of 
$O(8)\,$. From the result we identify the irreps which are constrained by the Bianchi 
identities $\nabla_{[\m}R_{\n\r]\s\l}\equiv 0\,$ and discard them. 
For example, if $(2,2)_{t.f.}$ corresponds to a tensor with the 
symmetries of the Weyl tensor (the subscript $t.f.$ stands for {\textit{trace-free}}), its first covariant derivative decomposes under $O(8)$ as 
$(1)\otimes (2,2)_{t.f.}\cong (3,2)_{t.f.}\oplus (2,2,1)_{t.f.} \oplus (2,1)_{t.f.}\,$. 
Now, if  $(2,2)_{t.f.}$ {\textit{is}} the Weyl tensor, the irrep $(2,2,1)_{t.f.}$ identically 
vanishes due to the Bianchi identities $\nabla_{[\m}R_{\n\r]\s\l}\equiv 0\,$.

At the end of this process, a set $S^i$ of $O(8)\,$ irreps
corresponds to every tensor 
$\nabla_{\!\a_1\ldots\a_i}C\,$, $i=0,\ldots,4$. 
As far as the quadratic terms in $A_8$ are concerned, the product of two tensors 
$(\nabla_{\a_1}\ldots\nabla_{\a_i}C)~(\nabla_{\a_1}\ldots\nabla_{\a_j}C)\,$, 
$\;i,j=0\ldots 4\,,\; i+j=4$ gives rise to $m$ independent scalars upon contraction if 
the set $S^i\cap S^j$ consists of $m$ elements. The simplest example is 
$S^4\cap S^1=\{ (2,2)_{t.f.}  \}\,$. Accordingly, there is one scalar $S$ obtained 
from the contraction of $C^{\a}_{~\,\b\g\eps}$ with (the double trace of) 
$\nabla_{(\a_1\ldots\a_4}C^{\b}_{~\,\g\d)\eps}\,$.    
The first term on the r.h.s. of  (\ref{ansatz8}) belongs to the same class as $S$ because 
it differs from the latter by terms in $J^k\,$, $k<6\,$.
As far as the cubic terms are concerned,  
$(\nabla_{\a_1}\ldots\nabla_{\a_i}C)(\nabla_{\a_1}\ldots\nabla_{\a_j}C)$
$(\nabla_{\a_1}\ldots\nabla_{\a_k}C)\,$, $\,i+j+k=2\,$, one must first take the tensor product 
of the $O(8)\,$-irreps corresponding to the first two tensors and then compare the resulting 
set of irreps $S^{i\otimes j}$ with $S^k\,$.    
\vspace*{.1cm}

{\bfseries{Remark :}} 
The problem of finding the number of locally independent 
scalars which can be built from polynomials in (covariant derivatives
of) the Weyl tensor is addressed in Appendix D of
\cite{Fulling}. We are not able to use these results as a check,
since unfortunately, some statements 
in Appendix D of \cite{Fulling} are erroneous. 
For example, at order 6, it is stated in \cite{Fulling} that there is only one 
scalar of the type ``$\nabla C\; \nabla C\,$''. This is actually not the case, since the 
following two scalars are locally independent : 
\begin{eqnarray}
A_6 &=& \nabla_{\a}C_{\b\m\n\r} \nabla^{\a}C^{\b\m\n\r}\,,
\nn \\
B_6 &=& \nabla^{\a}C_{\a\m\n\r} \nabla_{\b}C^{\b\m\n\r}\,.\nn   
\end{eqnarray}
One way to prove the independence of these two scalars is to expand them in 
the basis  
${\cal{R}}^0_{\{1\;1\}}\equiv\{\mathbf{e_1},\mathbf{e_2},\mathbf{e_3},\mathbf{e_4}\}$
$=\{ \nabla_{\a}R \nabla^{\a}R\,$,
$\nabla_{\a}R_{\b\c}\nabla^{\a}R^{\b\c}\,$, $\nabla_{\a}R_{\b\c}\nabla^{\c}R^{\b\a}\,$, 
$\nabla_{\a}R_{\b\c\d\eps}\nabla^{\a}R^{\b\c\d\eps}\}$ given in Appendix B of
the same 
 \cite{Fulling}. Explicitely, we find  
\begin{eqnarray}
A_6&=&\frac{2}{(D-2)(D-1)}\;\mathbf{e_1}-\frac{4}{(D-2)}\;\mathbf{e_2}+\mathbf{e_4}\,,
\nn \\
B_6&=&\frac{(D-3)^2}{2(D-2)^2(D-1)}\;\mathbf{e_1}+
\frac{2(D-3)^2}{(D-2)^2}\;(\mathbf{e_2}-\mathbf{e_3})\,.\nn     
\end{eqnarray}
Appendix D of \cite{Fulling} was obtained from the classification 
of polynomials in (covariant derivatives of) $R_{\a\b\c\d}\,$,
$R_{\a\b}$ and $R\,$, where $R_{\a\b}$ and $R$ were set 
to zero and where $R_{\a\b\c\d}$ was replaced by $C_{\a\b\c\d}$ throughout.
The problem is that this short-cut procedure is misleading because the Bianchi 
identities are  not carefully taken into account.  
For example, such a procedure gives ${\cal{C}}^0_{\{1\;1\}}$
$=\{\nabla_{\a}C_{\b\c\d\eps}\nabla^{\a}C^{\b\c\d\eps}\}\,$. 
However, if we make an invertible change of coordinates in ${\cal{R}}^0_{\{1\;1\}}$ and 
chooses the basis $\{\mathbf{e_1},\mathbf{e_2},2(\mathbf{e_2}-\mathbf{e_3}),\mathbf{e_4}\}$ instead, 
the contracted Bianchi identities ``create'' extra Riemann tensors : $2(\mathbf{e_2}-\mathbf{e_3})\equiv$ $\nabla^{\m}R_{\m\b\a\c}\nabla_{\n}R^{\n\b\a\c}\,$. 
Starting from the new basis, the ``short-cut procedure'' would now give a 
correct two-dimensional basis ${\cal{C}}^0_{\{1\;1\}}=\{ A_6,B_6 \}\,$.   
    
\vspace{0.5cm}

\subsection{Classification Theorem}
\label{theo}

Using the results of the previous sections, we may summarize our main
result about the classification of $D=8$ local conformal invariants 
in the form of a theorem. 

We obtain these invariants as follows.
We multiply our basis Ansatz $A_8\,$ given in (\ref{ansatz8}) 
by $\sqrt{-g}$ and take the 
Weyl transformation of the resulting density $\ca_8$ using  
(\ref{variaW}). Then we express every tensor 
$W^{\Delta_i}$ in terms of covariant derivatives of the Riemann tensor, the Ricci 
tensor and the scalar curvature. Using {\textit{FORM}} \cite{vermaseren} and 
{\textit{Ricci}} \cite{Lee}, the latter tensors are then decomposed in irreducible representations of $GL(8)\,$, so that the redundancies due to 
the differential Bianchi identities are eliminated. The redundancies due to 
the cyclic identity $R_{\a\b\c\d}+R_{\a\c\d\b}+R_{\a\d\b\c}\equiv 0$ are also 
taken into account. 
Finally, the condition $\d_{\s}\ca_8=0$ amounts to solving a system of linear equations 
for the eighteen
 coefficients entering the expression of $A_8\,$, which we do.  
The result of these operations is given by the 

\newpage
{\bfseries{Theorem [Classification of local Weyl-invariant scalar densities in $D=8$] 
}}{\textit{
Besides the seven Weyl invariants of the type $\sqrt{-g}\;C\,C\,C\,C$ 
given in \cite{Fulling}, 
there exist five Weyl-invariant scalar densities in dimension $D=8\,$,  }}
\begin{eqnarray}
\ci_j=\sqrt{-g}\,I_j\,,\quad j=1,\ldots, 5\,.   \nn
\end{eqnarray}
{\textit{The first one starts with the quadratic term 
$I_1=C^{\m\n\r\s}\Box^{2}C_{\m\n\r\s}+\cdots$  
whereas the other four are at least cubic in the Riemann tensor. 
Decomposing the scalars $I_j\,$, $j=1,\ldots,5$ in the natural 18-dimensional 
basis for Weyl-invariants given in  (\ref{ansatz8}) above, we have  }}   
{
\begin{eqnarray}
        \mathbf{I_1} &=& (1,48/25,2,42/125,9/10,3/5,96/125,74/25,208/5,
        \nonumber \\
        &&~~~~\quad\quad ~~~~~~~~~~~~~~~
        -8,16/5,-144/25,-104/5,0,0,-88/25,0,0)\,,
   \nonumber \\ 
   \mathbf{I_2} &=& (0,0,0,0,0,0,0,1,0,0,5,0,0,5,0,12/5,0,0)\,,
  \nonumber \\ 
   \mathbf{I_3} &=& (0,0,0,0,0,0,0,1,0,-20,0,-48/5,0,0,0,0,0,-20)\,,
   \nonumber \\ 
   \mathbf{I_4} &=& (0,0,0,0,0,0,0,1,10,-5/6,-5/24,4/5,-4,0,0,-7/5,5,0)\,,
   \nonumber \\ 
   \mathbf{I_5} &=& (0,0,0,0,0,0,0,1,8,-2/3,5/6,24/25,-16/5,0,-16/5,-12/25,0,0)\,.
\nonumber
\end{eqnarray}
}
\noindent This theorem is the main result of our paper.
For completeness, we give hereafter the seven quartic 
invariants taken from \cite{Fulling}, 
\begin{eqnarray}
        \ci_6 &=& \sqrt{-g}\; (C^{\p\q\r\s}C_{\p\q\r\s})^2\,,
        \nn \\
        \ci_7 &=& \sqrt{-g}\; C^{\p\q\r\s}C_{\p\q\r}^{~~~~\t}C^{\u\v\w}_{~~~~\s}C_{\u\v\w\t}\,,
        \nn \\
        \ci_8 &=& \sqrt{-g}\; C^{\p\q\r\s}C_{\p\q}^{~~\t\u}C_{\t\u}^{~~\v\w}C_{\r\s\v\w}\,,
        \nn \\
        \ci_9 &=& \sqrt{-g}\; C^{\p\q\r\s}C_{\p\q}^{~~\t\u}C_{\r\t}^{~~\v\w}C_{\s\u\v\w}\,,
        \nn \\
        \ci_{10} &=& \sqrt{-g}\; C^{\p\q\r\s}C_{\p\q}^{~~\t\u}C_{\r~\t}^{~\v~\w}C_{\s\u\v\w}\,,
        \nn \\
        \ci_{11} &=& \sqrt{-g}\; C^{\p\q\r\s}C_{\p~\r}^{~\t~\u}C_{\t~\u}^{~\v~\w}C_{\q\v\s\w}\,,
        \nn \\
        \ci_{12} &=& \sqrt{-g}\; C^{\p\q\r\s}C_{\p~\r}^{~\t~\u}C_{\t~\q}^{~\v~\w}C_{\u\v\s\w}\,.\nn
\end{eqnarray}
\vspace*{.1cm}
\noindent The eight-dimensional type B Weyl anomalies thus have the general structure 
\begin{eqnarray}
        \ca_8 = \sum_{j=1}^{12}\,a_j\;\int d^8 x \; \s \; \ci_j\,,\nn
\end{eqnarray}
where the model-dependent constants $a_j$ are 
determined within a given renormalization scheme.


\newpage

\noindent{\large{\bf Acknowledgements}}
\vspace*{.6cm}

\noindent N.~B.~is grateful to Marc Henneaux for having drawn his attention to the 
problem and for stimulating remarks.
We would like to thank Glenn Barnich, Helga Baum, 
Xavier Bekaert, Francis Dolan, Hugh Osborn,  
Christiane Schomblond, Christian Stahn and Stefan Theisen for useful discussions.  
N.~B.~is supported by a Wiener-Anspach grant (Belgium).
J.~E.~is supported by the Deutsche Forschungsgemeinschaft (DFG) within the Emmy
Noether programme, grant ER301/1-3.

\vspace{1cm}

{\small {\it Note added in proof.} It would be interesting to compare the
  method presented here with a suitable generalization of \cite{Deser}.}

\vspace{1cm}


\appendix
\section{Appendix}
\subsection{Decomposition of the Jet space}
\label{jetspace}
%
In this section, we decompose the jet space of the Riemann tensor in terms of the jet 
space of the Weyl tensor plus completely symmetrized covariant derivatives of the 
tensor $K_{\a\b}\,$. 

\vspace*{.2cm}
The expression (\ref{weyl2}) corresponds to a decomposition 
of the Riemann tensor under the group $O(D)$ (actually this decomposition is not irreducible because $K_{\a\b}$ is not traceless):   
\begin{center}
\begin{picture}(300,30)(0,-15)
(\ref{weyl2})
\put(12,0){$\Leftrightarrow$}
   \multiframe(35,3)(10.5,0){2}(10,10){}{}
   \multiframe(35,-7.5)(10.5,0){2}(10,10) {}{}
   \put(65,0){$\simeq$}
   \multiframe(84,3)(10.5,0){2}(10,10){}{}
   \multiframe(84,-7.5)(10.5,0){2}(10,10) {}{}
   \put(107,-8.5){\scriptsize{\mbox{t.f.}}}
   \put(123,0){$\bigoplus$}
   \multiframe(145,-2)(10.5,0){2}(10,10){}{}
\footnotesize
 \put(35,-20){$R_{\a\b\g\d}$}
 \put(84,-20){$C_{\a\b\g\d}$}
 \put(145,-20){$K_{\a\b}$}
\normalsize
\end{picture}
\end{center}
The subscripts ``t.f.'' indicates that the corresponding tensor
is trace-free.
The tensors $C_{\a\b\g\d}$ and $K_{\a\b}$ are two independent components
of $R_{\a\b\g\d}$, there exists no algebraic relation between 
$C_{\a\b\g\d}$ and $K_{\a\b}$.

At next order in the covariant derivatives of the Riemann tensor 
we have the following decomposition : 
\begin{center}
\begin{picture}(28,0)(3,5)
\multiframe(10,3)(10.5,0){1}(10,10){}
\put(10,-15){$\nabla_{\l}$}
\end{picture}
$\bigotimes$
\begin{picture}(28,15)(3,0)
\multiframe(10,3)(10.5,0){2}(10,10){}{}
\multiframe(10,-7.5)(10.5,0){2}(10,10) {}{}
\put(10,-20){$R_{\a\b\g\d}$}
\end{picture}
~~~$\simeq$
\begin{picture}(28,0)(3,5)
\multiframe(10,3)(10.5,0){1}(10,10){}
\put(10,-15){$\nabla_{\l}$}
\end{picture}
$\bigotimes$ 
\begin{picture}(28,0)(5,0)
\multiframe(10,3)(10.5,0){2}(10,10){}{}
\multiframe(10,-7.5)(10.5,0){2}(10,10) {}{}
\put(33,-7.5){\footnotesize{\mbox{t.f.}}}
\put(10,-20){$C_{\a\b\g\d}$}
\end{picture}
~~~$\bigoplus$~~~
\begin{picture}(45,0)(5,5)
\multiframe(10,3)(10.5,0){3}(10,10){}{}{}
\put(10,-15){$\nabla_{(\l}K_{\a\b)}$}
\end{picture}
$\bigoplus$
\begin{picture}(38,15)(3,0)
\multiframe(10,3)(10.5,0){2}(10,10){}{}
\multiframe(10,-7.5)(10.5,0){1}(10,10) {}
\put(10,-20){$ \tilde{C}_{\b\g\d}$}
\put(34,-7.5){\footnotesize{\mbox{t.f.}}}
\end{picture} \hspace{1ex} .
\end{center}
\vspace*{30pt}

The last Young diagram in the upper decomposition represents the Cotton tensor
$\tilde{C}_{\b\g\d} \equiv 2 \nabla_{[\delta}K_{\g]\b} $.
Actually this diagram is ``contained" in the following tableau\footnote{Of course,  
this holds only when $D>3$, since the Weyl tensor identically vanishes when $D=3\,$.} : 
\begin{center}
\begin{picture}(28,0)(3,5)
\multiframe(10,3)(10.5,0){1}(10,10){}
\end{picture}  
$\bigotimes$
\begin{picture}(28,0)(3,0)
\multiframe(10,3)(10.5,0){2}(10,10){}{}
\multiframe(10,-7.5)(10.5,0){2}(10,10) {}{}
\put(33,-7){\footnotesize{\mbox{t.f.}}}
\end{picture}~~~~~.
\end{center}
Indeed, if we decompose every Young tableau with respect to irreducible representations 
of $O(D)$, we see that 
\begin{center}
\begin{picture}(28,0)(3,5)
\multiframe(10,3)(10.5,0){1}(10,10){}
\end{picture}  
$\bigotimes$
\begin{picture}(30,20)(3,0)
\multiframe(10,3)(10.5,0){2}(10,10){}{}
\multiframe(10,-7.5)(10.5,0){2}(10,10) {}{}
\put(33,-7){\footnotesize{\mbox{t.f.}}}
\end{picture}
~~~~~~$\simeq$~~~
\Big[
\begin{picture}(28,0)(3,5)
\multiframe(10,3)(10.5,0){1}(10,10){}
\end{picture}  
$\bigotimes$
\begin{picture}(30,20)(3,0)
\multiframe(10,3)(10.5,0){2}(10,10){}{}
\multiframe(10,-7.5)(10.5,0){2}(10,10) {}{}
\put(33,-7){\footnotesize{\mbox{t.f.}}}
\end{picture}
~~~\Big]
\put(0,-7){\footnotesize{\mbox{t.f.}}}
~~~~~~$\bigoplus$~~
\begin{picture}(38,15)(3,0)
\multiframe(10,3)(10.5,0){2}(10,10){}{}
\multiframe(10,-7.5)(10.5,0){1}(10,10) {}
\put(34,-7.5){\footnotesize{\mbox{t.f.}}}
\end{picture} 
\end{center}
where the tracefree part of $\nabla_{\a} C_{\b\g\d\e}$ is 
\be
(\nabla_{\a} C_{\b\g\d\e})_{t.f.}=\nabla_{\a} C_{\b\g\d\e}+\frac{2(D-3)}{D}
\left[ g_{\a[\b}  \tilde{C}_{\g]\d\e}+g_{\a[\d}  \tilde{C}_{\e]\b\g}
\right].
\ee
In other words, if one takes the trace operation into account, 
 there appear algebraic relations between 
$\nabla_{\a} C_{\b\g\d\e}$ and the Cotton tensor : the latter is the trace of the former 
\be
 {\tilde{C}}_{\b\g\d}=\frac{-1}{D-3} \nabla_{\a} C^{\a}_{~\b\g\d}
                  =\frac{-1}{D-3} g^{\l\a} \nabla_{\l} C_{\a\b\g\d}.
\ee
At order $1$ in the derivatives of the Riemann tensor, we thus have the linearly 
independent tensors  $\nabla_{\a} C_{\b\g\d\e}$ and $\nabla_{(\l}K_{\a\b)}\,$.
Diagramatically, we represent this by : 
\begin{center}
  \begin{picture}(28,5)(3,5)
\multiframe(10,3)(10.5,0){1}(10,10){}
\put(10,-15){\footnotesize $\nabla_{\a}$}
\end{picture}  
$\bigotimes$
  \begin{picture}(38,15)(3,0)
\multiframe(10,3)(10.5,0){2}(10,10){}{}
\multiframe(10,-7.5)(10.5,0){2}(10,10) {}{}
\put(10,-20){\footnotesize $R_{\m\n\r\s}$}
\end{picture} 
$\simeq$
  \begin{picture}(28,0)(3,5)
\multiframe(10,3)(10.5,0){1}(10,10){}
\put(10,-15){\footnotesize $\nabla_{\a}$}
\end{picture} 
$\bigotimes$ 
\begin{picture}(28,0)(5,0)
\multiframe(10,3)(10.5,0){2}(10,10){}{}
\multiframe(10,-7.5)(10.5,0){2}(10,10) {}{}
\put(33,-7){\footnotesize {\mbox{t.f.}}}
\put(10,-20){\footnotesize $C_{\m\n\r\s}$}
\end{picture}
~~~~~$\bigoplus$~~
\begin{picture}(45,0)(5,5)
\multiframe(10,3)(10.5,0){3}(10,10){}{}{}
\put(10,-15){\footnotesize $\nabla_{(\a}K_{\b\g)}$}
\end{picture} .
\end{center}
\vspace*{20pt}
At order two in the covariant derivatives of the Riemann tensor, we use the previous 
decomposition to which we apply one more covariant derivative. Besides the second covariant derivative 
of the Weyl tensor, we obtain also the tensor $\nabla_{\m}\nabla_{(\a}K_{\b\g)}\,$. 
This tensor, when decomposed into irreducible representations of $GL(D)$, gives rise to 
a completely symmetric tensor $\nabla_{(\m}\nabla_{\a}K_{\b\g)}$ plus 
the tensor obtained by antisymmetrizing $\nabla_{\m}\nabla_{(\a}K_{\b\g)}$ with 
respect to the indices $\m$ and $\a\,$. The latter component is equivalent to the 
covariant derivative of the Cotton tensor, up to terms of lower order in the covariant 
derivatives of the Riemann tensor. 
We thus have the following decomposition in independent components :
\begin{center}
\begin{picture}(22,5)(3,5)
\multiframe(10,3)(10.5,0){1}(10,10){}
\end{picture}  
$\bigotimes$
  \begin{picture}(15,5)(9,5)
\multiframe(10,3)(10.5,0){1}(10,10){}
\end{picture} 
$\bigotimes$ 
  \begin{picture}(32,0)(9,0)
\multiframe(10,3)(10.5,0){2}(10,10){}{}
\multiframe(10,-7.5)(10.5,0){2}(10,10) {}{}
\end{picture} 
$\simeq$
  \begin{picture}(22,5)(5,5)
\multiframe(10,3)(10.5,0){1}(10,10){}
\end{picture}  
$\bigotimes$
  \begin{picture}(15,5)(9,5)
\multiframe(10,3)(10.5,0){1}(10,10){}
\end{picture} 
$\bigotimes$ 
\begin{picture}(35,0)(9,0)
\multiframe(10,3)(10.5,0){2}(10,10){}{}
\multiframe(10,-7.5)(10.5,0){2}(10,10) {}{}
\put(33,-10){t.f.}
\end{picture}
$\bigoplus$
  \begin{picture}(22,5)(3,5)
\multiframe(10,3)(10.5,0){4}(10,10){}{}{}{}
\put(-252,-15){\footnotesize $\nabla_{\m}$}
\put(-215,-15){\footnotesize $\nabla_{\n}$}
\put(-173,-15){\footnotesize $R_{\a\b\g\d}$}
\put(-128,-15){\footnotesize $\nabla_{\m}$}
\put(-88,-15){\footnotesize $\nabla_{\n}$}
\put(-53,-15){\footnotesize $C_{\a\b\g\d}$}
\put(13,-15){\footnotesize $\nabla_{(\a}K_{\b\g)}$}

\end{picture} 
\end{center}
\vspace*{20pt}

The same kind of decomposition appears at each order. 

\subsection{The representation matrices}
\label{T}

We give the expression for the representation matrices 
${\big[{\mathbf{T}}^{\m}\big]}^{\Delta_{i+1}}_{~~\Delta_{i}}\,$, 
$i=0,1,2,3\,$. We use the notation 
$\d^{\a_1\ldots\a_n}_{\b_1\,\ldots\b_n}\equiv \d_{\b_1}^{\a_1}\cdots \d_{\b_n}^{\a_n}\,$.
The tensor ${\cal{E}}_{\s\,\g\,\d\,\eps}^{\l\g'\d'\eps'}$ is a projector 
on tensors with the symmetry of the Weyl tensor (given explicitely in  
\cite{EO,Johanna}).
\begin{eqnarray}
        [\mathbf{T}^{\m}]_{\a_1~\g\d\eps\vert\bar{\b}}
        ^{~~\b~~~~~\bar{\g}\bar{\d}\bar{\eps}} = g^{\b\s}{\cal{E}}_{\s\,\g\,\d\,\eps}
        ^{\l\g'\d'\eps'}
        \big[  -2\d_{\a\l}^{a\r}-4\d_{\l\a}^{a\r}+4g^{a\r}g_{\l\a}\big]\,
        {\cal{E}}_{\r\,\g'\d'\eps'}^{\k\bar{\g}\;\bar{\d}\;\,\bar{\eps}}\;g_{\k\bar{\b}}\,.
\nn
\end{eqnarray}
\bqn
[\mathbf{T}^\m]_{\a_2\a_1~\g\d\eps\vert ~\;\bar{\b}}
^{~~~~~\b ~~~\a_1'~ \bar{\g}\bar{\d} \bar{\eps}}&=&
g^{\b\s}{\cal{E}}_{\s\,\g\,\d\,\eps}^{\r\g'\d'\eps'}
\left[ -6 \d^{\;\,\m\;\,\a_1'\;\s'}_{(\a_2\a_1)\r}-8\d^{\,\;\a_1'\s'\;\m}_{(\a_2\a_1)\r}
\right.
\nonumber \\
   &+& \left. 8g^{\m \s'}g_{\r(\a_1}\d^{\a_1'}_{\a_2)}+g_{\a_1\a_2}g^{\a_1'\m}\d^{\s'}_{\r}
\right]
{\cal{E}}_{\s'\,\g'\d'\eps'}^{\k\;\bar{\g}\;\bar{\d}\;\,\bar{\eps}}\;g_{\k\bar{\b}}
\,.\nn
\eqn
\bqn
{[\mathbf{T}^\m]}_{\a_3\a_2\a_1~\g\d\eps\vert ~~~~~\bar{\b}}
^{~~~~~~~\b~~~~\a_2'\a_1'~\bar{\g}\bar{\d}\bar{\eps}}
  &=&g^{\b\s}
{\cal{E}}_{\s\,\g\,\d\,\eps}^{\r\g'\d'\eps'}\left[  - 8 \d^{\,\a_2'\m~\a_1'\s'}
_{(\!\a_3\!\a_2\!)\a_1\:\r} - 3 \d^{\a_2'\m\;\a_1'\,\s'}_{\a_3\a_1\a_2\r}
 - \d^{\m\;\a_2'\,\a_1'\,\s'}_{\a_1\a_2\a_3\r}\right.
\nn \\
\!&-&\! 4 \d^{\a_2'\a_1'\s'\m}_{\a_3\a_2\a_1\r} - 
 8 \d^{~\!\a_2'\s'\,\a_1'\,\m}_{(\!\a_3\!\a_2\!)\a_1\:\r} 
+ 4 g^{\m\s'}g_{\r\a_1}\d^{\a_2'\a_1'}_{\a_3\a_2}
+8 g^{\m\s'}g_{\r(\a_3}\d^{\a_2'\a_1'}_{\a_2)\a_1}
\nn \\
\!&+&\! \left. 2g^{\a_1'\m}g_{\a_1(\a_2}\d^{\a_2'\s'}_{\a_3)\r} + 
 g_{\a_2\a_3}g^{\m\a_2'}\d^{\a_1'\s'}_{\a_1\r} \right]  
 {\cal{E}}_{\s'\,\g'\d'\eps'}^{\k\;\bar{\g}\;\bar{\d}\;\,\bar{\eps}}\;
 g_{\k\bar{\b}}\,.
 \nn
\eqn
\begin{eqnarray}
{[\mathbf{T}^\m]}_{\a_3\a_2\a_1\a~\g\d\eps\vert ~~~~~~~\bar{\b}}
^{~~~~~\,\,~~~\b~~~~\a_2'\a_1'\a'~\bar{\g}\bar{\d}\bar{\eps}}&=&
\d^{\a_2'}_{\a_3}       {[\mathbf{T}^\m]}_{\a_2\a_1\a~\g\d\eps\vert ~~~~~\bar{\b}}
^{~~~~~~\b~~~~\a_1'\a'~\bar{\g}\bar{\d}\bar{\eps}}+
{[\mathbf{C}^\m]}_{\a_3\a_2\a_1\a~\g\d\eps\vert ~~~~~~~\bar{\b}}
^{~~~~~\,\,~~~\b~~~~\a_2'\a_1'\a'~\bar{\g}\bar{\d}\bar{\eps}}\,,\nonumber\\
{[\mathbf{C}^\m]}_{\a_3\a_2\a_1\a~\g\d\eps\vert ~~~~~~~\bar{\b}}
^{~~~~~\,\,~~~\b~~~~\a_2'\a_1'\a'~\bar{\g}\bar{\d}\bar{\eps}}&=&
g^{\b\s}{\cal{E}}_{\s\,\g\,\d\,\eps}^{\r\g'\d'\eps'}\left[
-5\d_{\a_3\a_2\a_1\a\,\l}^{\,\m\;\a_2'\a_1'\a'\s'}-\d_{\a_2\a_3\a_1\a\,\l}^{\,\m\;\a_2'\a_1'\a'\s'}-\d_{\a_1\a_2\a_3\a\,\l}^{\,\m\;\a_2'\a_1'\a'\s'}\right.
 \nonumber \\
 &-& \d_{\a\a_2\a_1\a_3\,\l}^{\,\m\;\a_2'\a_1'\a'\s'} + 
 g_{\a_3\a_2}g^{\m\a_2'}\d_{\a_1\a\,\l}^{\a_1'\a'\s'} +
 g_{\a_3\a_1}g^{\m\a_1'}\d_{\a_2\a\,\l}^{\a_2'\a'\s'} \nonumber \\
 &+& 
\left. g_{\a_3\a}g^{\m\a'}\d_{\a_2\a_1\,\l}^{\a_2'\a_1'\s'} 
 + 4 g_{\a_3\l}g^{\m\s'}\d_{\a_2\a_1\a}^{\a_2'\a_1'\a'} - 
 4 \d_{\l\a_2\a_1\a\a_3}^{\,\m\;\a_2'\a_1'\a'\s'}\right]
 {\cal{E}}_{\s'\,\g'\d'\eps'}^{\k\;\bar{\g}\;\bar{\d}\;\,\bar{\eps}}\;g_{\k\bar{\b}}\nn\,.
\end{eqnarray}


\end{document}